\documentclass[aps,prd,twocolumn,letterpaper]{revtex4}
\usepackage[dvips]{graphicx}

\def\ie{{\em i.e.\ }}
\def\eg{{\em e.g.\ }}
\def\etal{{\em et al.\ }}
\def\p{\partial}
\def\d{{\rm d}}

\def\dd#1#2{\frac{\d #1}{\d #2}}
\def\pp#1#2{\frac{\p #1}{\p #2}} 
\def\rmmat#1{{\hbox{\rm #1}}}
\def\rmscr#1{\rmmat{\scriptsize #1}}
\def\acr#1{{\hat a}^\dag_{#1}}
\def\aan#1{{\hat a}_{#1}}

\begin{document}

\title{Quantum Mechanical Fluctuations at the End of Inflation}
\author{Jeremy S. Heyl\footnote{Canada Research Chair}}
\address{Department of Physics and Astronomy, 
University of British Columbia \\ 
6224 Agricultural Road, Vancouver, British Columbia, Canada, V6T 1Z1}

\date{\today}

\begin{abstract}
  During the inflationary phase of the early universe, quantum
  fluctuations in the vacuum generate particles as they stretch beyond
  the Hubble length.  These fluctuations are thought to result in the
  density fluctuations and gravitational radiation that we can try to
  observe today. It is possible to calculate the quantum-mechanical
  evolution of these fluctuations during inflation and the subsequent
  expansion of the universe until the present day.  The present
  calculation of this evolution directly exposes the particle creation
  during accelerated expansion and while a fluctuation is larger than
  the Hubble length. Because all fluctuations regardless of their
  scale today began as the vacuum state in the early universe, the
  current quantum mechanical state of fluctuations is correlated on
  different scales and in different directions.
\end{abstract} 
\pacs{PACS numbers: 98.80.Cq, 98.80.Qc }
\maketitle

\section{Introduction}

When one thinks of macroscopic manifestations of quantum mechanics one
usually thinks of phenomena such as superconductivity and lasers that
have crept at least somewhat into our everyday lives.  However, if
there was an inflationary epoch in the early universe, quantum
mechanical fluctuations and correlations determined the large-scale
structure of the universe, and galaxies and superclusters are 
quantum mechanics writ upon the largest scales of our Universe.

During inflation a slowly rolling scalar field (the inflaton) drives
an exponential expansion of spacetime.  This rapid expansion addresses
several cosmological issues.  It ensures that space is very close
to flat and that topological defects that form before and early during
inflation will be exceedingly rare today.  Finally the quantum
mechanical evolution of the perturbations in a spatially uniform
inflaton field (or other fields) provide the seeds for structure
formation in the recent universe and for the fluctuations of the
microwave background.

There are several equivalent ways to calculate the evolution of
fluctuations in scalar fields during the approximately de~Sitter
inflationary period.  Possibly the most straightforward way is
to exploit the fact that the equation describing the evolution of the
field operator is the same as that describing the classical
evolution of a scalar field \cite{Lidd00}; one can start with a
reasonable approximation for the wavefunction of the de~Sitter vacuum
and evolve it forward through the de~Sitter stage.  At the onset of
radiation domination, one can use the method of Bogolubov
coefficients \cite{1988PhRvD..37.2078A} to ensure continuity of the
field and its time derivative.

Studies of how fluctuations in the early universe lose coherence have
followed two main paths.  The first looks at a quantum field coupled
to the environment: for example, a thermal bath
\cite{1989PhRvD..40.1071U} or the backreaction of pair production of
the quantum state of the universe \cite{1995PhRvD..52.6770C}.  The
second path is to examine how the appearance of decoherence develops
in de~Sitter space without any decohering interaction.  Work to
understand how initially quantum fluctuations begin to act classically
as they are stretched outside the Hubble length moved forward dramatically
with the work of Guth and Pi \cite{1982PhRvL..49.1110G}.  They found
that long after a mode passes through the Hubble length it follows a classical
Gaussian probability distribution.  Several people have addressed the
quantum-to-classical transition in terms of squeezed states
\cite{1990PhRvD..42.3413G,1993PhRvL..70.2371G,1994PhRvD..50.4807A}.
Albrecht \etal \cite{1994PhRvD..50.4807A} argued that although the
the concept of squeezed states is useful in understanding the
evolution of density fluctuations, it is entirely equivalent to more
traditional methods and does not provide any new physical consequences.

This paper presents an alternative method that explicitly follows the
particle creation.  It is perhaps closest in spirit to the recent work
of Miji\'c \cite{1998PhRvD..57.2138M} that follows the evolution of
particle number as a mode is stretched beyond the Hubble length.
Rather than treat the evolution of the field in the standard manner
using a field operator that follows the de~Sitter vacuum, we can use a
operator that allows the wavefunction is expanded as a superposition
of multi-particle states as observed by a small comoving detector.
The entire wavefunction is calculated as a function of time, rather
than the expectation value of a particular operator (typically the
field operator).  This method is computationally intensive, so it
cannot be applied to modes that spend many Hubble times larger than
the Hubble length.  However, it offers new insights into how density
fluctuations approach the classical limit.

In particular one can calculate the particle number using both
techniques and they agree, but one can also construct any operator
after the wavefunction has been calculated and determine its
expectation value or the probabilities of the various outcomes of a
particular measurement.  \S\ref{sec:evol-scal-field} will derive the
Hamiltonian of a scalar field in an homogeneous, isotropic universe and
define the field operators for the de~Sitter vacuum
(\S\ref{sec:vacuum-picture}) and the vacuum of a radiation-dominated
universe (\S\ref{sec:particle-picture}).  The evolution of the
wavefunction is described in \S\ref{sec:radi-domin-spac} and its
initial conditions in \S\ref{sec:initial-conditions}.

The results of the calculation (\S\ref{sec:results}) fall in two
categories.  First, we examine the robust of the calculations and
compare with the results from the standard treatments in
\S\ref{sec:number-particles}.  We find that this treatment predicts
the same amount of particle production as in the standard treatment
(\S\ref{sec:vacuum-picture} and \eg \cite{1997gr.qc.....7062F}) and
that the number of particles follows a thermal distribution.  This
highly excited Fock state behaves quasi-classically; however, the state
of the field is still pure and furthermore the phases of the various
multiparticle states are correlated (\S\ref{sec:prim-corr}).
\S\ref{sec:discussion} speculates on the possibility of observing
these correlations and future work.

\section{The Evolution of Scalar-Field Modes}
\label{sec:evol-scal-field}

We will follow the evolution of perturbations of a uniform real scalar
field in a homogeneous and isotropic spacetime with flat spatial
sections.  The metric is 
\begin{equation}
ds^2 = a^2(\tau) \left ( d\tau^2 - d {\bf x}^2 \right ).
\end{equation}
where $a(\tau)$ is the scale factor.   
The action for the scalar field ($\phi$) is given by
\begin{equation}
S = 
\int d^4 x \sqrt{-g} \left [ \frac{1}{2} \partial_\mu \phi \partial^\mu \phi -
  V(\phi) \right ].
\end{equation}
where we have taken $\hbar=c=1$ and $\sqrt{-g}=a^4(\tau)$.   We neglect the back reaction of the
perturbations on the metric.  From the action we would like to find
the Hamiltonian and determine the quantum mechanical evolution of the fluctuations.

Now let's take $V(\phi)=\frac{1}{2} \left ( m^2 + \xi R \right )\phi^2 + V_0$
where the scalar field may have a non-minimal coupling to gravity
through the Ricci scalar $R$ \cite{1997PThPh..98.1063K}. 
With these we obtain
%
%
\begin{eqnarray}
S &=& 
\frac{1}{2} \int d^4 x \sqrt{-g} \left [ \partial_\mu  \phi
  \partial^\mu  \phi - (m^2 + \xi R) \phi^2 \right ]  \\
&=& 
\frac{1}{2} \int d^4 x a^2  \biggr [ 
\left ( \pp{\phi}{\tau} \right )^2 - \left (\nabla  \phi
\right)^2 \nonumber \\
& & ~~~~~~ - a^2 \left ( m^2 + \frac{6\xi}{a^3} \pp{^2 a}{\tau^2}  \right )\phi^2 \biggr ] 
\end{eqnarray}
where we have dropped the constant term $V_0$.
Let's make the substitution $u=a \phi$ to try to absorb the
factor of $a^2(\tau)$ into the fields.
\begin{eqnarray}
S &=& 
\frac{1}{2} \int d^4 x  \Biggr [ 
\left ( a \pp{(u/a)}{\tau} \right )^2 - \left (\nabla u\right)^2 
\nonumber \\
& & ~~~~~~
 - a^2 \left ( m^2 + \frac{6\xi}{a^3} \pp{^2 a}{\tau^2}\right ) u^2
 \Biggr ]  \\
&=& \frac{1}{2} \int d^4 x  \Biggr [ 
\left ( \pp{u}{\tau} - \pp{a}{\tau} \frac{u}{a}\right )^2 - \left (\nabla u\right)^2 
\nonumber \\
& & ~~~~~~
  - a^2 \left ( m^2 + \frac{6\xi}{a^3} \pp{^2 a}{\tau^2}\right ) u^2
 \Biggr ]  \\
&=& \frac{1}{2} \int d^4 x  \Biggr [ 
\left ( \pp{u}{\tau} \right )^2 + \dd{}{\tau} \left ( -\frac{u^2}{a}
  \pp{a}{\tau} \right ) + \frac{1}{a}\pp{^2 a}{\tau^2} u^2 
\nonumber \\
& & ~~~
  - \left (\nabla u\right)^2  -
 a^2 \left ( m^2 + \frac{6\xi}{a^3} \pp{^2 a}{\tau^2}\right ) u^2
 \Biggr ]  
\\ 
&=& \frac{1}{2} \int d^4 x  \Biggr [ 
\left ( \pp{u}{\tau} \right )^2 - \left (\nabla u \right)^2  
\nonumber \\
& & ~~~~~~
 -
\left ( a^2 m^2 + \frac{6\xi-1}{a} \pp{^2 a}{\tau^2} \right ) u^2 \Biggr ] 
\end{eqnarray}
where to get the final result we have dropped a term in the integrand 
equal to a total derivative with respect to the conformal time.

The field $u({\bf x},\tau)$ can be expressed in terms of Fourier
modes,
\begin{equation}
u({\bf x},\tau) = \frac{1}{(2 \pi)^{3/2}} \int d^3 k u_{\bf k}(\tau)
e^{i{\bf k}\cdot {\bf r}}
\end{equation}
where $u_{\bf -k}(\tau) = u^*_{\bf k}(\tau)$ to ensure that 
the field $u$ is real.  This yields
\begin{equation}
S = \frac{1}{2} \int d\tau d^3 k  \left [ 
\left | \pp{u_{\bf k}}{\tau} \right |^2 - \left ( k^2 + 
 m^2_\rmscr{eff}  \right ) \left |u_{\bf k}\right | ^2 \right ] 
\end{equation}
where ${\bf k}$ is the comoving momentum of a mode.  

Now we have the action for a related scalar field $u$ in a flat
space-time with a negative contribution to the square
of the mass of the mode \cite{1992PhR...215..203M,Lidd00}.

The value of the mass depends on the evolution of the background
spacetime (\ie $a(\tau)$).  If the dominant energy
density in the universe is characterised by a pressure that is
proportional to the energy density ($P=w\rho$), The scale factor evolves as
\begin{equation}
a(\tau) \propto \left \{ \begin{array}{ll}
e^{(aH)\tau} & \rmmat{if}~w = -\frac{1}{3} \\
\tau^{2/(3w+1)} & \rmmat{otherwise}
\end{array}
\right .
\end{equation}
where $H$ is the Hubble parameter, $(\partial a/\partial
\tau)/a^2$. For $w=-1/3$, the product of the scale factor and the
Hubble parameter is constant.  The effective mass is 
\begin{equation}
m^2_\rmscr{eff} = -2 \frac{Q}{\tau^2}
\label{eq:1}
\end{equation}
with
\begin{equation}
Q \equiv \frac{1}{(1+3w)^2} \left [ 
\left ( 1 - 3w \right ) \left (1 - 6 \xi\right ) - 2 \frac{m^2}{H^2}
\right ]
\end{equation}
if $w\neq -1/3$.  The special case $w=-1/3$ yields
\begin{equation}
m^2_\rmscr{eff} = (a H)^2 \left ( 6\xi-1 + \frac{m^2}{H^2} \right )
\end{equation}

The Hamiltonian of the field is given by
\begin{equation}
H = \frac{1}{2} \int  d^3 k  \left [ 
\left | \pp{u_{\bf k}}{\tau} \right |^2 + \left ( k^2 + 
 m^2_\rmscr{eff}  \right ) \left | u_{\bf k} \right|^2 \right ] .
\end{equation}
To look at the quantum mechanics of this Hamiltonian one can take 
one of several routes, depending on the definition of the field operator
${\hat u}_{\bf k}$.  In general it is \cite{Lidd00}
\begin{equation}
{\hat u}_{\bf k} = g(k, \tau) \aan{\bf k} + 
g^*(k,\tau) \acr{-\bf k}
\end{equation}
where $\acr{\bf k}$ and $\aan{\bf k}$ are creation and
annihilation operators that satisfy the following standard commutator
relations
\begin{equation}
\big [ \acr{\bf k}, \acr{\bf k'} \big ] = 
        \big [ \aan{\bf k}, \aan{\bf k'} \big ] = 0,
        \big [ \aan{\bf k}, \acr{\bf k'} \big ] = 
                        \delta^{(3)} ({\bf k} - {\bf k'}).
\label{eq:2}
\end{equation}
There is some flexibility in choosing the functions $g(k,\tau)$ -- or
equivalently defining the states upon which the creation and annihilation
operators act.

Specifically one can choose $g(k,\tau)$ so that the Hamiltonian
operator commutes with the particle number operator, $\acr{\bf
  k}\aan{\bf k}$, and all of the time dependence is carried by the
field operator or one can choose $g(k,\tau)$ so that the number of
particles changes. The first option is tractable in closed form if $Q$
is constant or piecewise constant in $\tau$; that is if $m\neq 0$ and
$w=-1$ or $m=0$ and $w$ is constant or piecewise constant.

\subsection{Vacuum Picture} 
\label{sec:vacuum-picture}

In the first case, the Hamiltonian operator is given by
\begin{equation}
{\hat H} = \frac{1}{2} \int \d^3 k\  k 
\left (
\acr{\bf k} \aan{\bf k} + \aan{\bf k} \acr{\bf k} \right ) .
\end{equation}
The Hamiltonian commutes with the particle number operator, so the
universe remains in the vacuum state, $\left | 0 \right >$.  If the
Hamiltonian takes this form, the functions $g(k,\tau)$ must satisfy
the following differential equation
\begin{equation}
\pp{^2 g(k,\tau)}{\tau^2} + \left ( k^2 - 2 \frac{Q}{\tau^2} \right ) g(k,\tau)
= 0
\label{eq:3}
\end{equation}
If $Q$ is constant we have
\begin{equation}
g(k,\tau) = \sqrt{\tau} \left [ A J_\nu (k\tau) + B Y_\nu(k\tau)
  \right ] 
\label{eq:4}
\end{equation}
where $\nu = \sqrt{1+8Q}/2$.  Two solutions of particular importance
are for $Q=1$ for $w=0,-1$ and $\xi=m=0$ and $Q=0$ for $w=1/3$ and 
$\xi=m=0$.  We have
\begin{equation}
g(k,\tau) = \frac{1}{\sqrt{2k}} e^{-ik\tau}
\label{eq:5}
\end{equation}
for $Q=0$ and
\begin{equation}
g(k,\tau) = \frac{1}{\sqrt{2k}} \left ( k\tau - i \right ) \frac{e^{-ik\tau}}{k\tau}
\label{eq:6}
\end{equation}
for $Q=1$ \cite{Lidd00}.  For $w=-1$ (de~Sitter space) one can use the
general solution Eq.~(\ref{eq:4}) when $m\neq 0$ and $\xi\neq 0$, and
during radiation domination ($w=\frac{1}{3}$), 
one can find a general solution in terms
of Whittaker functions even when $m\neq 0$ and $\xi\neq 0$.   In this
paper the focus will be massless scalar fields, \ie perturbations to
the inflaton field and gravitational waves that can be modelled as
two independent massless scalar fields \cite{1988PhRvD..37.2078A};
therefore, the solutions given by Eq.~(\ref{eq:5}) and~(\ref{eq:6})
will suffice for the radiation-dominated and de~Sitter  phases, respectively.

In general if $Q$ is only piecewise constant, one must insist that the
function $g(k,\tau)$ and its first derivative are continuous through
the transition, because Eq.~(\ref{eq:3}) is second order.  This
determines the values of $A$ and $B$ in Eq.~(\ref{eq:4}) after the
transition (\ie the Bogolubov transformation). 

In particular if we look at the transition from vacuum domination to
radiation domination, we find that Eq.~(\ref{eq:6}) applies during
vacuum domination and using the general solution, Eq~(\ref{eq:4}), we
find that 
\begin{eqnarray}
g(k,\tau) &=& \frac{1}{\sqrt{2k}} \frac{x_\rmscr{max}^2}{2}
\biggr [  
\left ( 2 x_\rmscr{max}^{-2} - 2 i x_\rmscr{max}^{-1} - 1 \right )
 e^{-i k \tau} \nonumber \\
& & ~~~+  e^{-2 i / x_\rmscr{max}} e^{i k \tau}
\biggr ]
\label{eq:7}
\end{eqnarray}
during the subsequent radiation dominated epoch 
where $x_\rmscr{max}=1/(k \tau_\rmscr{RH})=a_\rmscr{RH} H/k$, the
ratio of the Hubble parameter to physical wavenumber at the end 
of inflation.

From Eq.~(\ref{eq:7}) we can read off the number of particles that are
created in a particular mode from an initial vacuum state; it is
simply given by the squared of the coefficient of the negative
frequency term (see for example Eq.~1.26 of
\cite{1997gr.qc.....7062F}),
\begin{equation}
\langle N \rangle = \frac{1}{4} \left ( \frac{a_\rmscr{RH}H}{k} \right )^4
\label{eq:9}
\end{equation}

\subsection{Particle Picture}
\label{sec:particle-picture}

In this second technique one chooses a particular function for
$g(k,\tau)$ and sticks with it for the entire calculation.
Specifically, we choose Eq.~(\ref{eq:5}).  This has two advantages.
First, this choice of $g(k,\tau)$ is appropriate for radiation
domination, \ie the state of the universe after inflation -- so the
number of particles in a particular state remains constant after
inflation.  Second, this $g(k,\tau)$ is appropriate for modes whose
physical wavelengths are much smaller than the Hubble length, \ie the
situation in the distant past and the distant future (today).  The
particle number operator, $\acr{\bf k}\aan{\bf k}$, does not commute
with the Hamiltonian during the de~Sitter epoch, and its expectation
value gives the number of particles that we would expect to observe
with a particle detector looking for excitations small compared to the
present-day scale of the universe.  Although there is considerable
freedom in choosing the functions $g(k,\tau)$ due to the vacuum
ambiguity that appears in quantum field theory in curved spacetime,
this particular choice of $g(k,\tau)$ gives the canonical quantization
of a scalar field in a flat spacetime, in particular the locally flat,
inertial frame of a comoving detector, imparting some conceptual
advantages of this picture over that standard picture outlined in
\S\ref{sec:vacuum-picture}.

Because the Lagrangian for a scalar field in a Robertson-Walker
spacetime evolves the same way as a scalar field in a flat spacetime
but with a time-dependent mass, it is quite natural to express the
Hamiltonian for the system as the sum of the flat space Hamiltonian
\cite{Grei96} and time-dependent portion
\begin{eqnarray}
{\hat H} &=& \frac{1}{2} \int \d^3 k\  \Biggr [ \left ( k
- \frac{Q}{\tau^2 k} 
 \right )\left (
\acr{\bf k} \aan{\bf k} + \aan{\bf k} \acr{\bf k} \right )  \nonumber
\\ 
& &  - \frac{Q}{\tau^2 k} 
\left ( 
\aan{-{\bf k}} \aan{\bf k} e^{-2 i k \tau}
+ \acr{-{\bf k}} \acr{\bf k} e^{2 i k \tau}
  \right )
\Biggr ].
\label{eq:8}
\end{eqnarray}
$\aan{\bf k}$ annihilates the vacuum of a radiation-dominated universe, 
$\aan{\bf k} \left |0\right > = 0$.

The advantage of this technique is that it connects an initial state
(the vacuum) to the final state in the basis in which we observe it.
Because the final state is effectively the large-scale structure of
the universe, following the quantum mechanical evolution of the scalar
field becomes exponentially cumbersome for modes much larger than the
Hubble scale.  Nevertheless, even the short-term evolution of modes as
they expand beyond the Hubble scale provides new insights.

\subsection{Radiation-Dominated Spacetime Fock Space}
\label{sec:radi-domin-spac}

The Hamiltonian consists of a particle-conserving portion (which is
the sum of the radiation-dominated result and a time-dependent correction)
and a component which creates and destroys particles.  The entire 
Hamiltonian commutes with the total momentum, so it is natural to 
examine how the Hamiltonian acts on states with zero total momentum.
Let us take the following sum of states
\begin{eqnarray}
\left | \psi \right > &=& \sum_{n=0}^\infty B_n(\tau) \
\frac{\big ( \acr{\bf k'} \big )^n
\big ( \acr{\bf -k'} \big )^n}{n!\left [ \aan{\bf k'}, \acr{\bf k'} \right ]^n} 
 \left |\ 0\ \right > \\
&=& 
\sum_{n=0}^\infty B_n(\tau) \
\left | n, -{\bf k'}; n, {\bf k'} \right >,
\label{eq:10}
\end{eqnarray}
where we have assumed that the vacuum is normalised. 
 
$|B_n(\tau)|^2$ is the probability of a comoving observer in the
distant future detecting $n$ particles with comoving 
momentum ${\bf k}'$ and $n$ particles with momentum $-{\bf k}'$ if 
inflation ended at a conformal time $\tau$.
This yields
\begin{eqnarray}
{\hat H} \left | \psi \right > &=& 
\sum_{n=0}^\infty B_n(\tau) \Biggr [
 \left ( k - \frac{1}{\tau^2 k} \right )
( 2 n + Z )  \left | n, -{\bf k}; n, {\bf k} \right >
\nonumber \\
& & - \frac{1}{\tau^2 k} \biggr ( 
n \left | n-1, -{\bf k}; n-1, {\bf k} \right >  e^{-2 i k \tau}
 \\
& & ~~~ + (n+1) \left | n+1, -{\bf k}; n+1, {\bf k} \right > e^{2 i k \tau}
\biggr ) \nonumber
\label{eq:13}
\end{eqnarray} 
where $Z=\big [ \aan{\bf k}, \acr{\bf k} \big ] = 
                        \delta^{(3)} ({\bf k} - {\bf k})$ 
is an infinite constant related to the vacuum energy of the
system.  Its value is independent of the number of particles in the
states.

We have replaced ${\bf k}'$ with ${\bf k}$ after performing the
integral over all momenta in Eq.~(\ref{eq:8}).  The integral with our
choice of basis functions is manifestly even in ${\bf k}$ so the
factor of one-half vanishes.  Furthermore, the Hamiltonian yields a
solution separable in momentum-space, so we can solve for the
evolution of each value of comoving momentum separately.  As shown in
the subsequent paragraphs, by solving this for a single value of the
comoving momentum in de~Sitter space one obtains solutions useful over
a range of momenta.

The time evolution of the coefficients $B_n(\tau)$ is given by
the value of 
\begin{eqnarray}
\left <n, -{\bf k}; n, {\bf k} \right |{\hat H} \left | \psi \right >
&=& i \frac{\d B_n(\tau)}{\d\tau} \\
&=& \Biggr [ 
 \left ( k - \frac{Q}{\tau^2 k} \right )
( 2 n + Z ) B_n(\tau) \nonumber \\
& & ~
- \frac{Q}{\tau^2 k} \biggr ( 
B_{n-1} (\tau) n e^{2 i k \tau}  \\
& & ~~~~ + B_{n+1} (\tau) (n+1) e^{-2 i k \tau} \biggr )  \Biggr ] .
\nonumber
\label{eq:14}
\end{eqnarray}
Since ${\hat H}$ evolves a state forward in time we obtain,
\begin{eqnarray}
i \dd{A_n(x)}{x} &=& -Q \Biggr [
m A_{n-1}(x) e^{-2 i \gamma(x)/x} \nonumber \\*
& & ~
+ (n+1) A_{n+1}(x) e^{ 2 i \gamma(x)/x}
 \Biggr ]
\label{eq:evolution}
\end{eqnarray}
where we changed variables to $x=-1/(k\tau)$ and absorbed an infinite
phase into the definition of $A_n$,
\begin{eqnarray}
A_n (x) &=& e^{-i (2 n + Z) (\gamma(x)-1)/x } B_n(x) 
\label{eq:15}\\
\gamma(\tau) &=&  2 + Q x^2.
\label{eq:16}
\end{eqnarray}
The variable $x$ equals $aH/k$ during the de~Sitter phase and $-aH/k$
during radiation domination. The value of $x$ increases with time during both
these phases.  The physical momentum $p$ is given by $p=k/a$, so
$p/H=1/|x|$.  The value of $Q$ during the vacuum-energy-dominated
phase ($w=-1$) is unity.  During radiation domination ($w=1/3$), in
the massless case $Q$ vanishes, so the coefficients $A_n$ are constant
during the radiation-dominated epoch (see Eq.~\ref{eq:evolution}).

It is important to keep in mind that the functional relationship
between the effective mass and the conformal time dictated both the
change of variables and the (infinite) phase shift of the
wavefunctions. In a situation where $Q$ is not constant with respect
to $\tau$, a different set of substitutions are necessary.  In general
we must integrate up the time-dependence of the effective mass
(Eq.~\ref{eq:15}) to renormalise the infinite zero-point energy of the
field.  This could be done numerically of course.

Because $A_n(\tau)$ differs from $B_n$ only by a phase, $|A_n(\tau)|^2$
still is the probability of a comoving observer detecting $n$
particles with comoving momentum ${\bf k}$ and $n$ particles with
comoving momentum $-{\bf k}$ during radiation domination or the
distant future if inflation ends at conformal time $\tau$.
Furthermore, because the differential equation only depends on the
product of $k \tau$, the solution to the equation gives the
wavefunction in the Fock space of a radiation-dominated spacetime
for a range of comoving momenta.

\subsection{Initial Conditions}
\label{sec:initial-conditions}

Initially, well before the mode crosses outside of the Hubble length,
$|A_0(\tau \rightarrow -\infty)|^2=1$ and there are no particles.
During inflation, the mode expands relative to the Hubble length.  At
the end of inflation, $\tau=-\tau_\rmscr{RH}$, the universe reheats
and the energy in the inflaton field decays into relativistic
particles.  After reheating we can no longer follow the quantum
fluctuations of the inflaton field itself; however, any correlations
in the inflaton field should also be present in the resulting
perturbation to the density of relativistic matter.  The quantum
fluctuations of a fluid follow equations similar that of the scalar
field \cite{1992PhR...215..203M}, specifically the value of $Q$ is the
same for scalar fields and fluids.  We can also follow the evolution
of the perturbations in a second scalar field which does not couple to
the inflaton as they cross back within the Hubble length.  The
perturbations induced by inflation on the metric itself can be
decomposed into two minimally coupled scalar fields
\cite{1988PhRvD..37.2078A}, so the primordial spectrum of
gravitational radiation falls into this category and furthermore these
fluctuations remain in the linear regime until the present epoch.

For simplicity, the vacuum and radiation-dominated regimes are pasted
together so that after the end of inflation $\tau=\tau_\rmscr{RH}$,
\ie we assume that reheating is instantaneous and that the universe
does not expand during this time.  The fact that inflation ends
suddenly results in production of particles at all momenta.
Specifically, the total energy of the particles produced diverges in
the high-frequency limit; however, the sudden approximation is only
valid for modes whose periods are much greater than the timescale for
the transition.  Higher frequency modes will evolve adiabatically so
their final state will be the vacuum state.  We also neglect any
nonlinear couplings that may develop during reheating.  This is
probably reasonable for tensor fluctuations but non-linearities may
play an important role for density fluctuations and this is an avenue
for further work.

The conformal time, $\tau$, continues to increase until the present
day.  The number of states occupied as well as the mean number of
particles per mode increases dramatically with the time that a
particular mode spends outside of the Hubble length during an inflationary
period.  On the other hand, during radiation domination, the
wavefunctions evolve only trivially, the coefficients $A_n$ are
constant.

\section{Results}
\label{sec:results}

As discussed in the previous paragraph, the natural initial condition
for the simulation when a mode is much smaller than the Hubble length is
$\left |\psi \right > = \left | 0, {\bf k} ; 0, -{\bf k} \right >$.
We simulated the evolution of a massless real scalar field using this
initial condition and Eq.~(\ref{eq:evolution}).
To assess the accuracy of the simulations, the initial states $\left |
1, {\bf k} ; 1, -{\bf k} \right >$ and $\left | 2, {\bf k} ; 2, -{\bf
k} \right >$ are also evolved forward. 
The inner product between these states and the initial vacuum state
vanishes initially.
\begin{figure}
\centerline{\includegraphics[height=3.5in]{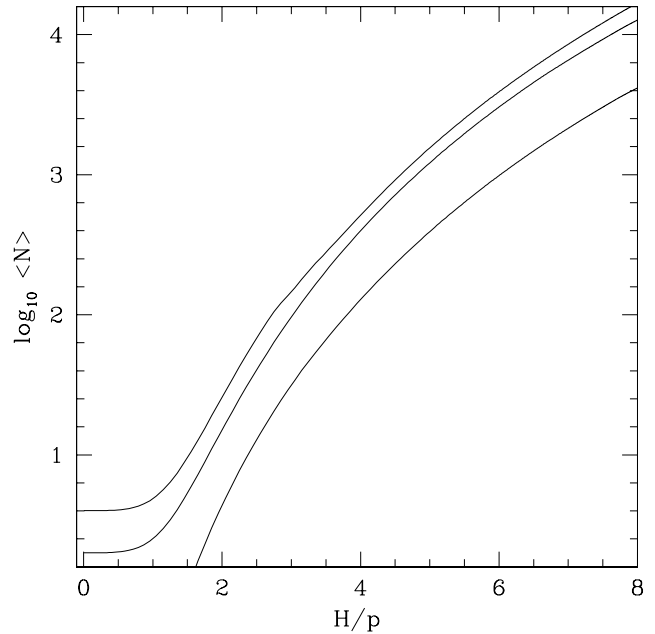} }
\smallskip 
\caption{Evolution of the expectation value of the particle number before
and soon after Hubble length exit.  The upper curve is for a four-particle
initial state, the middle curve is a two-particle initial state and
the lower curve is a vacuum initial state.}
\label{fig:multi}
\end{figure}

\subsection{Number of Particles}
\label{sec:number-particles}

\begin{figure}
\centerline{\includegraphics[height=3.5in]{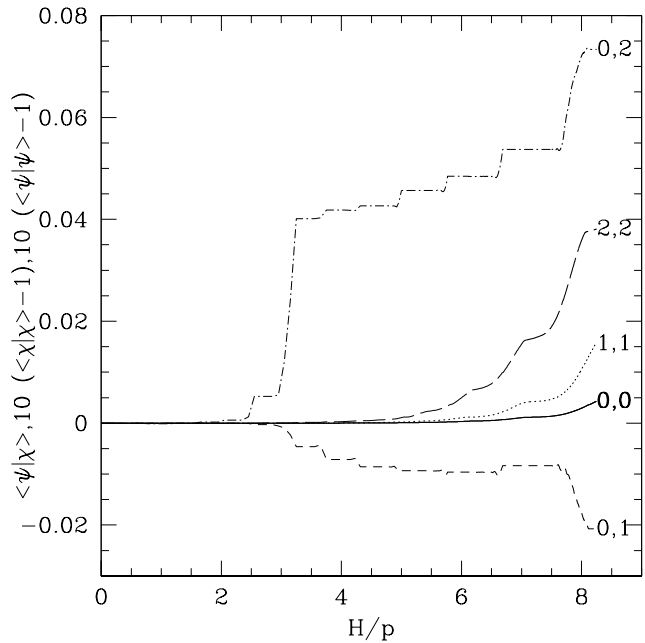}}
\smallskip 
\caption{Evolution of the scalar product between the various initial
  states. For clarity the value of $10 \left (\sum A_n^* A_n -
  1\right)$ is depicted as well as the cross products.  If the
  simulation were free of numerical errors, these values would all vanish.}
\label{fig:dp}
\end{figure}

As Fig.~\ref{fig:multi} depicts, each initial state results in
thousands of particles being produced over a wide range of comoving momenta.
The rate of particle production is dramatically larger than that found
by Miji\'c \cite{1998PhRvD..57.2138M}.  Also these calculations do not
exhibit the discontinuity that Miji\'c found at $H/p=1/\sqrt{2}$ when
the mass of a mode becomes imaginary.  Although the pair production is
prodigious and the final states have little overlap with the initial
states, the accumulated numerical error estimated by the scalar
products and the conservation of the norm of the states is modest
(Fig.~\ref{fig:dp}).  The accumulated error increases in a series of
periodic steps rather than continuously.  Each of these steps
corresponds approximately to a conformal time where $2 \gamma(x) = n
\pi$ where $n$ is a negative integer (see Eq.~\ref{eq:16}).

For all three sets of initial conditions, the number of particles
increases dramatically for the modes that have exited through the
Hubble length; however, in no case does the particle number increase
exponentially with the comoving momentum; this would be indicated by a
straight line in Fig.~\ref{fig:multi}.

\begin{figure}
\centerline{\includegraphics[height=3.5in]{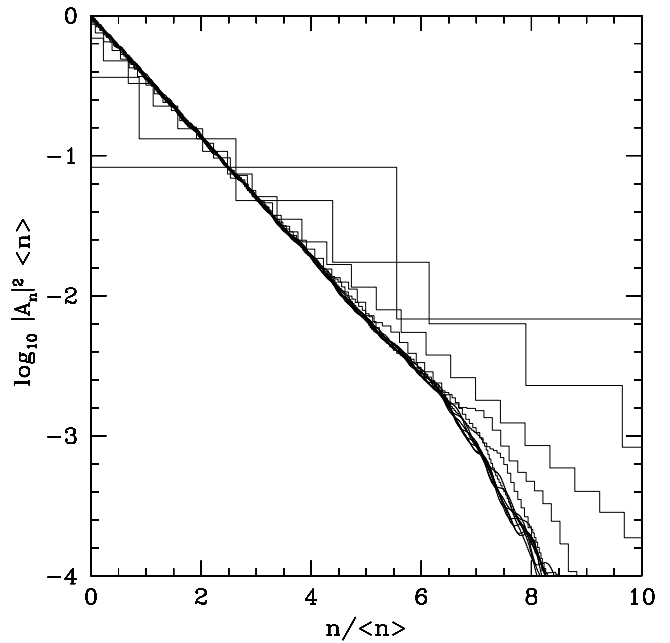}}
\smallskip 
\caption{The curves trace the final values of $|A_n|^2$ as a function
  of $n/\langle n \rangle$.  The curves are normalised by the
  expectation value of $n$.
  Starting at the lower right corner of the panel and moving left, the
  curves trace modes with the following maximum values of $x\equiv H/p,
  1.0, 1.5, 2.0 \ldots 8.0$. The distributions for $x>3$ are nearly identical.}
\label{fig:tailnorm}
\end{figure}

\begin{figure}
\centerline{\includegraphics[height=3.5in]{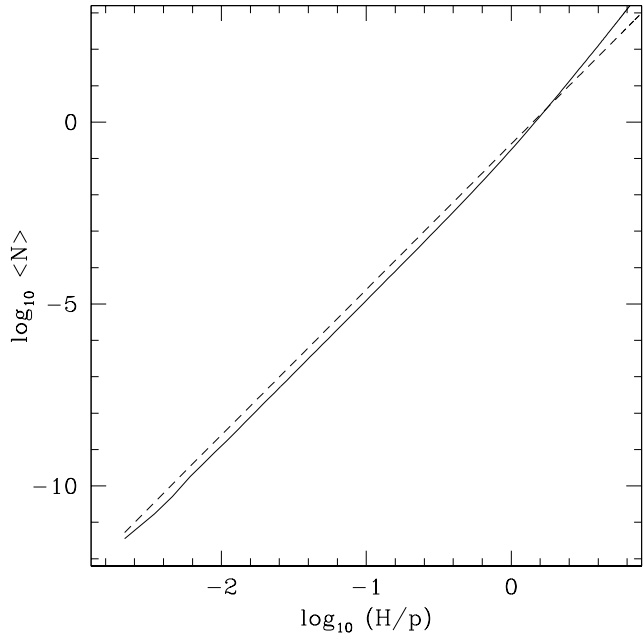}}

\caption{The solid line traces the calculated value of $\left <N
  \right >=\left <2 n \right>$ as a function the ratio of the physical
  momentum to the $H$ is the Hubble parameter at the end of inflaton
  in the particle picture.  The figure is similar to
  Fig.~\ref{fig:multi} but focuses on high frequency modes.  The
  dashed line shows the value given by Eq.~(\ref{eq:9}) in the vacuum
  picture.}
\label{fig:early}
\end{figure}

The simulations show that the occupation of the states is 
well described by a thermal distribution (Fig~\ref{fig:tailnorm}), 
Such a distribution may be characterised by a single parameter,
the expectation value of $n$.  The total number of particles is of
course $N=2n$.
Fig.~\ref{fig:early} depicts the expectation value of $2n$
as a function of the ratio of the physical momentum to the Hubble
parameter at the end of inflation.  
We have from Eq.~(\ref{eq:9}), the number of particles in a particular mode
\begin{equation}
\left <N \right > \approx \frac{1}{4} \left ( \frac{p}{H} \right )^{-4}
\end{equation}
Because the field theory for the field $u$ is defined for a comoving
volume, the physical energy density of particles produced within a
range of physical momenta $dp$ is
\begin{equation}
\dd{E}{V dp} \approx \frac{1}{8\pi^2} \frac{H^4}{p}
\end{equation}
so the total energy density of the modes diverges as $\ln
p_\rmscr{max}/p_\rmscr{min}$ where $p_\rmscr{max}\approx 1/\Delta t$
is some momentum cutoff, set by the duration of reheating
\cite{1997gr.qc.....7062F}.  If the product
of the physical momentum and the duration of reheating is much greater
than one, the value of $Q$ changes sufficiently slowly that the
initial vacuum oscillation will evolve adiabatically to the final
vacuum of the radiation dominated epoch.   The infrared momentum
cutoff is given approximately by the Hubble constant during reheating,
$p_\rmscr{min}\approx H$ \cite{1997gr.qc.....7062F}.

\subsection{Primordial Correlations}
\label{sec:prim-corr}

At some point the expansion of the universe becomes radiation
dominated.  If we assume that this transition is abrupt, we can
calculate the subsequent evolution of a mode during the period of
radiation domination as it shrinks relative to the Hubble length.  The
coefficients $A_n$ remain constant during this time because $Q=0$ (see
Eq.~\ref{eq:evolution}); in particular, the number of particles
remains constant.  When matter begins to dominate the energy density,
the coefficients $A_n$ again change with time as long as the mode lies
outside the Hubble length because now $Q\neq 0$.  The focus here will
be modes that only spend at most a few Hubble times outside the Hubble
length, so they will be well within the Hubble length at the onset of
matter domination, and the values of $A_n$ at reheating will
characterise the fluctuations today.

Even for modes that only spend a short time outside the Hubble length
the final states long after reheating contain a large number of
particles (Fig.~\ref{fig:multi}).  However, the partition function for
a given mode remains thermal.  Fig.~\ref{fig:tailnorm} shows that a
simple rescaling of the distributions by the mean number of particles
maps the final distributions into each other.

Although the partition function for the number of particles is thermal
for the final state of the fluctuations, the final wavefunctions for
different values of the comoving momentum are correlated.

%
%
%
%
%
%
%

\begin{figure}
\centerline{\includegraphics[height=3.5in]{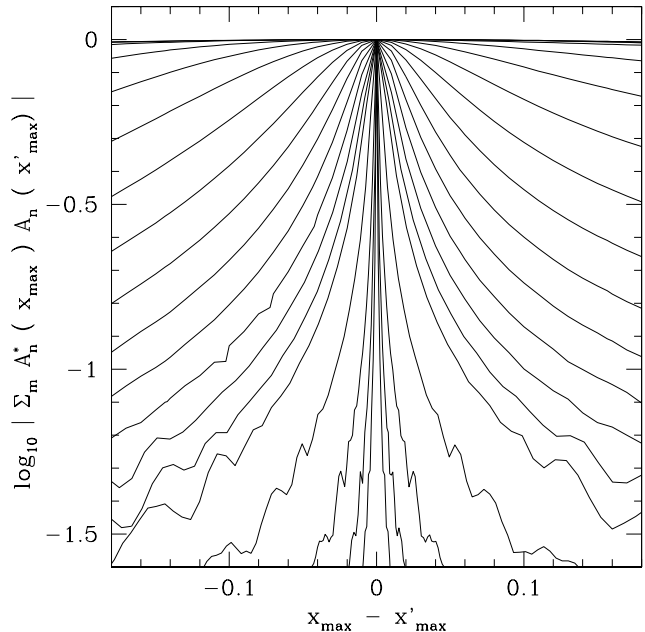}}
\smallskip 
\caption{The final value of $\sum_n A_n^* A'_n$ for different values of
  $x_\rmscr{max} = (H/p)_\rmscr{max}$.  Starting with the uppermost
  curve on the left $x_\rmscr{max}=1.6, 1.8, 2.0 \ldots 4.4,5,6,7,8$.
}
\label{fig:pairdp}
\end{figure}

The modes approach the classical regime when $x_\rmscr{max}$ gets larger
as Fig.~\ref{fig:pairdp} shows.  The figure depicts the scalar product
between the probability amplitudes of the final wavefunctions of modes
that have spent nearly the same amount of time outside the Hubble length or
have slightly different values of $k$.  We find that for modes that only
spend a short time outside the Hubble length the final state is quite
similar over a wide range of momentum.  The range over which the modes
correlate decreases as $x_\rmscr{max}^{-3}$ for $x_\rmscr{max}\gg 1$.
As $x_\rmscr{max}$ increases the correlation between the modes
diminishes.  An important fact to keep in mind is that although the
obvious correlations between the wavefunctions vanish as
$x_\rmscr{max}$ gets larger, the wavefunctions are still correlated
albeit more subtly; the phases of the probability amplitudes are not 
strictly random but pseudo-random, that is to say calculable.

Unfortunately, these correlated quantum phases are not directly
observable.  The wavefunction for each value of the comoving momentum
resides in a separate Hilbert space, so our observations of the
fluctuations on these small scales are drawn from the same underlying
distribution but are otherwise independent.  Only the most gross
properties of the distribution are correlated such as the probability
to find a certain number of particles in a particular state.  On the
other hand the correlated phases have a great pedagogical value in that
they verify that the production of fluctuations during inflation does
not require the destruction of information; it is strictly unitary.

\section{Sensitivity to the Details of Reheating}
\label{sec:sens-deta-rehe}

Although the simulations here assumed that reheating quickly converts
the potential energy of the scalar field to the kinetic energy of
relativistic particles, simulations of a transition to a
matter-dominated regime yielded very similar results.  The final
distributions are still thermal as in Fig.~\ref{fig:tailnorm} and the
correlations between the quantum states of modes with different values
of ${\bf k}$ are similar to Fig.~\ref{fig:pairdp} except that in the
case of a matter-dominated reheating, the correlations are typically
weaker for given values of ${\bf k}$ and ${\bf k}'$.

Of course, this is not a general demonstration that these results are
robust with respect to the details of reheating but it does indicate
that quantum correlations may persist to the present day.

\section{Discussion}
\label{sec:discussion}

The foregoing results show that fluctuations on scales of the
present-day universe that passed through the Hubble length near the
end of the inflationary epoch exhibit quantum mechanical correlations
that may belie their birth from the vacuum.  Today the comoving scale
of the Hubble length at the end of inflation is
\begin{equation}
\frac{1}{k_\rmscr{RH}} = 10^5 \left ( \frac{M}{10^{14}\rmmat{GeV}} \right )^{2/3}
\left ( \frac{T_\rmscr{RH}}{10^{10}\rmmat{GeV}} \right )^{1/3}
\rmmat{cm} 
\end{equation}
where $M^4$ is the vacuum energy associated with the inflaton field
and $T_\rmscr{RH}$ is the temperature of the universe at the end of
reheating.  The analysis has assumed that reheating is quick and
efficient \cite{1997PhRvD..56.3258K,2004PhDT.......414Z} so
$T_\rmscr{RH} \sim M$, yielding
\begin{equation}
\frac{1}{k_\rmscr{RH}} = 2 \times 10^6 \left (
\frac{M}{10^{14}\rmmat{GeV}} \right )^{1}
\rmmat{cm} 
\label{eq:21}
\end{equation}
The value of $x_\rmscr{RH}$ for the comoving scale $1/k_\rmscr{RH}$ is
simply one and $k = k_\rmscr{RH}/x_\rmscr{max}$.  Consequently
although the correlations are present on all scales, they are most
obvious on the comoving scale of the Hubble length at the end of inflation
(\ie really small scales).  On these small scales the density fluctuations
are well into the non-linear regime today but tensor fluctuations,
gravitational waves (GW), would still be a loyal tracer of these
correlations.

The expression given in Eq.~\ref{eq:21} is very uncertain.  A simple
way to quantify the chance that we might observe these correlations
directly is to calculate the number of $e-$foldings between the
time when our present Hubble scale equalled the Hubble scale during
inflation and the typical scale of future GW observatories probing inflation
(\eg the Big-Bang Observatory \cite{2005PhRvD..72h4001B}):
\begin{equation}
\Delta N = \ln \frac{0.1~\rmmat{Hz}}{H} \approx 38.
\end{equation}
Typically today's Hubble scale is assumed to pass out through the
Hubble length during inflation about 50$-$60 $e-$foldings \cite{Lidd00}
before the end, so the millihertz scale would pass through the 
Hubble length 12$-$22 $e-$foldings before the end.   However, the former 
number is highly uncertain.  For example, if inflation occurs at a
higher energy scale or if there is a epoch of late ``thermal
inflaton''
\cite{1996PhRvD..53.1784L,1995PhRvL..75..398D,1996NuPhB.458..291D},
the number of $e-$foldings for today's Hubble scale could
be as low as 25 \cite{Lidd00}.

Even if we eventually observe the fluctuations on the comoving scales
corresponding to the Hubble length at the end of reheating, one could
argue that the wavefunction only determines the probability of
measuring a particular amplitude and phase for the primordial
gravitational waves at a particular momentum, and one would need to
measure these fluctuations in several different realisations
(universes) to unearth the correlations.  

The value of these correlations lies in that they verify that the
production of fluctuations during inflation does not require the
destruction of information; it is strictly unitary.  Furthermore, one
can use the wavefunction to calculate the density matrix for
observations restricted to a portion of the spacetime (\eg within the
horizon \cite{2007arXiv0708.3353S}) and estimate the entanglement
entropy of the observable fluctuations.  The calculations can also be
easily extended to include non-linearities in the field either through
gravity or self-interactions.  Again these interactions generate
entanglement entropy when fluctuations on a single scale are
observed, giving the appearance of entropy production while the scale
field remains in a pure state.   Both of these calculations will be
addressed in a future paper.

\bigskip
\paragraph*{Acknowledgments}

The author would like to thank Avi Loeb and Bill Unruh for useful
discussions and Jim Zibin and Douglas Scott for comments on the paper.
He acknowledges support from NSERC. The calculations were performed on
computing infrastructure purchased with funds from the Canadian
Foundation for Innovation and the British Columbia Knowledge
Development Fund. This work made use of NASA's Astrophysics Data
System.

\bibliographystyle{prsty}
\bibliography{mine,gr,inflation,qed}

\end{document}